\documentclass[aps,prb,twocolumn,superscriptaddress,floatfix]{revtex4-2}
\usepackage{amsmath,amsfonts,amssymb,color}
\usepackage{amsthm}
\usepackage{braket}
\usepackage{graphicx}
\usepackage{xcolor}
\usepackage{dcolumn}
\usepackage{bm}
\usepackage{epstopdf}
\usepackage{epsfig}
\usepackage{environ}
\usepackage{pdfcomment}
\usepackage{float}
\usepackage[T1]{fontenc}
\usepackage[latin9]{inputenc}
\usepackage{setspace}
\usepackage{esint}

\begin{document}

\title{Coupled-wire construction of non-Abelian higher-order topological phases}
\author{Jiaxin Pan}
\affiliation{%
	College of Physics and Optoelectronic Engineering, Ocean University of China, Qingdao, China 266100
}
\author{Longwen Zhou}
\email{zhoulw13@u.nus.edu}
\affiliation{%
	College of Physics and Optoelectronic Engineering, Ocean University of China, Qingdao, China 266100
}
\affiliation{%
		Qingdao Key Laboratory of Advanced Optoelectronics, Qingdao, China 266100
}
\affiliation{%
	Engineering Research Center of Advanced Marine Physical Instruments and Equipment of MOE, Qingdao, China 266100
}
\date{\today}

\begin{abstract}

Non-Abelian topological charges (NATCs), characterized by their noncommutative algebra, offer a framework for describing multigap topological phases beyond conventional Abelian invariants. Although higher-order topological phases (HOTPs) host boundary states at corners or hinges, their characterization has largely relied on Abelian invariants such as winding and Chern numbers. Here, we propose a coupled-wire scheme for constructing non-Abelian HOTPs
and analyze in detail a two-dimensional model as its minimal realization. The resulting Hamiltonian supports hybridized corner modes, protected by parity-time reversal ($\mathcal{PT}$) plus sublattice symmetries, and described by a topological vector that unites a non-Abelian quaternion charge with an Abelian winding number. Corner states emerge only when both invariants are nontrivial, whereas weak topological edge states of non-Abelian origin arise when the quaternion charge is nontrivial, enriching the bulk-edge-corner correspondence. The system further exhibits topological phase transitions of both non-Abelian and Abelian characteristics, providing a unified platform that bridges these two distinct topological classes. Our work thus extends the study of HOTPs into non-Abelian regimes and suggests feasible experimental realizations in synthetic quantum matter.

\end{abstract}

\maketitle
\section{Introduction}

Topological matter has emerged as an active research area in quantum and condensed matter physics over the past decades \cite{REVTI1,REVTI2,REVTI3,REVTI5,REVTI6,REVTI7,REVTI4}. Principally, a topological phase is characterized by global order parameters, protected by spatial and internal symmetries, holds degenerate edge states associated with its bulk topology, and shows topologically quantized responses to external perturbations. Moreover, a topological transition between distinct phases of the same symmetry class does not accompany any symmetry breaking, making it go beyond the traditional Landau-Ginzburg-Wilson paradigm of phase transitions~\cite{REVTI8}. From the discovery of integer quantum Hall effects~\cite{QHE,TKNN} to theoretical and experimental progress on quantum anomalous Hall effects~\cite{AQHE1,AQHE2,AQHE3} and quantum spin Hall effects~\cite{QSHE1,QSHE2,QSHE3,QSHE4}, a general framework for understanding topological phases of matter is gradually established, which has profoundly reshaped our understanding of quantum matter and spurred continued research on topological materials~\cite{REVTI9,REVTI10,REVTI11}. More recently, the exploration of topological phenomena has been extended to non-Hermitian~\cite{REVNHT1,REVNHT3,REVNHT2,REVNHT4,REVNHT5} and periodically driven (Floquet) setups~\cite{REVFT,REVFT2,REVFT3,REVFT4,REVNHFT}, further extending the scope of the field to nonequilibirum and open systems. Building on these foundations, the classification of topological matter has been formulated within the Altland-Zirnbauer symmetry class. This so-called ``tenfold way'' provides a unified framework for characterizing all possible topological phases of free fermions in different spatial dimensions according to their time-reversal, particle-hole, and chiral symmetries~\cite{Tenfold}, laying the groundwork for later extensions to non-Hermitian~\cite{NHTClass1,NHTClass2,NHTClass3} and driven systems~\cite{FTClass,FTClass2,FTClass3}. Typically, the topological phases considered in these studies are described by $\mathbb{Z}$ or $\mathbb{Z}_2$ invariants, such as Chern and winding numbers. As these invariants are commutative and belong to Abelian groups, they could be regarded as Abelian topological charges.

\begin{figure*}
    \centering
    \includegraphics[width=0.9\linewidth]{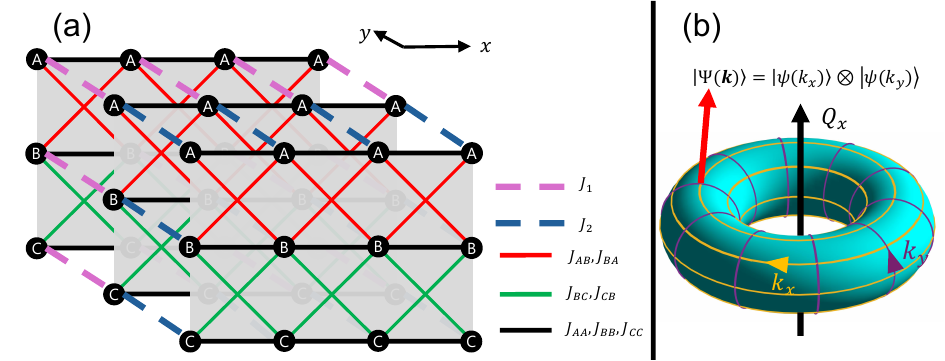}
    \caption{Illustration of the 2D lattice model and its Brillouin zone. (a) shows the model realized by stacking 1D trimer chains. Along $x$ direction, each unit cell has three sublattices with onsite potentials $S_{A}$, $S_{B}$, $S_{C}$. Couplings between neighboring sites are indicated by red and green lines, corresponding to $J_{AB(BA)}$ and $J_{BC(CB)}$, respectively, where we take $J_{AB}=J_{BA}= i u$ and $J_{BC}=J_{CB}= i v$. Couplings between the same kind of sublattices are shown as black bonds, representing $J_{AA}$, $J_{BB}$ and $J_{CC}$. Along $y$ direction, coupling strengths alternate between $J_{1}$ and $J_{2}$ in purple and blue dashed lines. (b) shows the 2D Brillouin zone on a torus.}
    \label{fig: lattice model}
\end{figure*}

Meanwhile, the concept of non-Abelian topological charge (NATC) has attracted increasing attention over the years. It provides a unified framework for describing noncommutative topological structures in systems with multiple bands/gaps and possessing ${\cal PT}$ or 
${\cal C}_{2}{\cal T}$ symmetry. NATC was originally utilized to characterize the biaxial nematic liquid crystal~\cite{BNLC1,BNLC3,BNLC2}, and recently been incorporated into the study of emerging topological semimetals and insulators \cite{NATS1,NATS2,NATS3,NATS4,NATS5,NATS6,NATS7,NATS8,NATS10,NATS9,NATS11,NATI1,NATI2,NATI3,NATI5,NATI4,NATI6,NATI7,NATI8,HONATI1,HONATI2,HONATI3,FNATI1,NHNATI,FNATI3,FNATI4,ZhouMobius2025} that fall outside the tenfold way of conventional symmetry classifications.
In non-Abelian topological semimetals, the NATC is used to characterize the knot and braiding structures of nodal lines or nodal points in different band gaps~\cite{NATS1}. 
Such distributed multigap degeneracies lead to complex topological configurations that cannot be captured by standard approaches based on Abelian geometric phases. In two-dimensional (2D) three-band systems, such gapless degeneracies can be characterized by the Euler class~\cite{NATS2}, whose non-Abelian topology has been experimentally observed~\cite{NATS3,NATS4,NATS5,NATS6,NATS7}.
In non-Abelian topological insulators (NATIs), the NATC can be used to distinguish different topological phases that correspond to distinct configurations of edge states located in different band gaps. This framework overcomes the limitations of Abelian geometric phases, such as the Zak or Berry phases, which could not identify all possible topological states and establish their bulk-edge correspondence. In particular, even when the underlying Bloch states of the system are all real vectors and the resulting Zak phase is trivial,
topological edge states of non-Abelian origins may still emerge under the open-boundary condition (OBC). Non-Abelian topological phases thus go beyond the conventional understanding of topological insulators and superconductors. Following rapid theoretical developments, experimental progress has also been made in realizing non-Abelian topological phases in metamaterials and quantum simulators, including transmission line networks~\cite{NATI1,NATI2}, photonics \cite{NATI3,NATI4,FNATI4}, acoustics \cite{NATI5,NATI6,FNATI3}, and ultracold atoms~\cite{NATI7}.

In the past decade, a unique class of topological phases was revealed, in which topological boundary states can emerge on ``higher-order surfaces'' (such as the corners or hinges of a cubic material) of the system. They are thus referred to as HOTPs. 
More precisely, a $d$-dimensional, $n$th-order topological phase ($1< n\leq d$) has symmetry-protected boundary states of dimensions $d-n$ \cite{HOTI1,HOTI2,HOTI3,HOTI6,HOTI5,HOTI8,HOTI9,HOTI7,HOTI10,HOTI11}.
In the early development of this concept, the
topological quadrupole insulator plays a key role, demonstrating the very existence of HOTPs and providing a rigorous theoretical foundation for their description~\cite{HOTI1}. 
Typically, HOTPs require crystalline symmetries~\cite{HOTI1,HOTI2,HOTI3,HOTI6,HOTI5}, such as rotation, mirror inversion, and other spatial symmetries for their protections.
Meanwhile, a correspondence was established between the existence of zero-dimensional corner states and the winding numbers of one-dimensional (1D) edges in sublattice-symmetric higher-order topological insulators (HOTIs)~\cite{HOTI6}, making it possible to realize HOTPs even without crystalline symmetries~\cite{HOTI7,HOTI8,HOTI9}.
Until now, HOTPs have been realized in different experimental settings~\cite{REVHOTI2,REVHOTI1,REVHOTI3} and considered very recently in non-Abelian regimes~\cite{HONATI1,HONATI2,HONATI3}.

In this work, we introduce a unique class of non-Abelian HOTPs, which are obtainable via staggered couplings of 1D non-Abelian topological chains along extra dimensions. Focusing on a 2D lattice model with hybridized non-Abelian and Abelian topology as illustrations, we reveal a rich set of non-Abelian second-order topological phases (SOTPs) with degenerate corner states, which are protected by ${\cal PT}$ and chiral (sublattice) symmetries. The resulting phase diagram of the system shows multiple phases and transitions. Each of them is characterized by a hybridized topological change, which contains a quaternion and an integer as its components. Further connections are established among these hybridized changes, topological corner states, and 1D non-Abelian edge bands, offering a unified perspective on the bulk-edge-corner correspondence. Our results thus extend the study of HOTPs to non-Abelian regimes and establish a connection between non-Abelian and Abelian topological matter in an individual setup. The rest of the paper is organized as follows.
In Sec.~\ref{sec: model and phase}, we introduce our theory, sketch the general idea for its implementation,
and construct our non-Abelian 2D model of SOTPs [see Fig.~\ref{fig: lattice model}(a) for an illustration of its lattice geometry].
In Sec.~\ref{sec: topological invariant}, hybridized non-Abelian topological charges are defined to characterize both the non-Abelian and Abelian topology of our system on an equal footing, leading to its bulk topological phase diagram.
In Sec.~\ref{sec: spectrum OBC}, we study topological boundary states of our system under OBC and reveal its bulk-edge-corner correspondence of non-Abelian characteristics.
The found topological phases host not only corner states but also weak non-Abelian edge bands, with both of them being captured by the hybridized charge we introduced.
In Sec.~\ref{sec: conclusion}, we summarize our results and discuss experimental prospects.
Further theoretical and numerical details are given in the Appendices \ref{app: Zak phase}--\ref{app: disorder}.

\section{Theory and model}\label{sec: model and phase}
In this section, we first introduce a theoretical framework to construct non-Abelian HOTPs. The general idea is to couple non-Abelian and/or Abelian topological lattices along different spatial dimensions, yielding a higher-dimensional system whose HOTPs are determined by the product topology of its lower-dimensional subsystems, as detailed in Sec.~\ref{sec:The}. The non-Abelian HOTPs thus obtained are referred to as pristine (hybridized) if none (some) of its constituting subsystems are characterized by Abelian topological charges. The minimal case following such a construction is given by hybridized non-Abelian SOTPs in two dimensions, whose lattice Hamiltonian is introduced and analyzed in Sec.~\ref{sec: Model}. More detailed treatments of the topology and phase transitions in our 2D minimal model are given in later sections.

\subsection{Theory}\label{sec:The}
In previous studies, the coupled-wire construction has been implemented to realize Abelian topological phases, such as 2D Su-Schrieffer-Heeger (SSH) type models~\cite{2DSSH1,2DSSH2,2DSSH3} and strongly interacting HOTPs~\cite{SIHOTP1,SIHOTP2,SIHOTP3,SIHOTP4}. This approach has also been extended to periodically driven systems~\cite{HOTI9,ZhouHOTP2,ZhouHOTP3,ZhouHOTP4,ZhouHOTP5}, enabling the exploration of Floquet HOTPs. Despite these advances, its applicability to systems hosting NATCs remains largely unexplored.
Here, we generalize the coupled-wire scheme for building a unique
class of non-Abelian HOTPs. We start with a set of Hamiltonians
$\{H_{x_{n}}|n=1,2,...,d-1,d\}$, where each $H_{x_{n}}$ describes
a 1D chain of free fermions along a separate dimension. Taking
the Kronecker sum among different $H_{x_{n}}$, we obtain the
lattice Hamiltonian of a $d$-dimensional system, i.e.,
\begin{equation}
H=H_{x_{1}}\oplus H_{x_{2}}\oplus\cdots\oplus H_{x_{d-1}}\oplus H_{x_{d}}.\label{eq:GeneralH}
\end{equation}
Here, $A\oplus B=A\otimes\mathbb{I}_{B}+\mathbb{I}_{A}\otimes B$
for any operators $A$ and $B$, with $\mathbb{I}_{A}$ and $\mathbb{I}_{B}$
being identities of the Hilbert spaces of $A$ and $B$. The system described
by $H$ could possess emergent HOTPs due to the interplay of lower-dimensional
topological states among its subsystems $H_{x_{n}}$. For example,
if each $H_{x_{n}}$ describes a topological insulator with a topologically
nontrivial edge state $|\psi_{x_{n}}\rangle$ of energy $E_{x_{n}}$,
i.e., $H_{x_{n}}|\psi_{x_{n}}\rangle=E_{x_{n}}|\psi_{x_{n}}\rangle$
for $n=1,...,d$, the composite system $H$ must have a localized
topological state $|\psi\rangle=|\psi_{x_{1}}\rangle\otimes|\psi_{x_{2}}\rangle\otimes\cdots\otimes|\psi_{x_{d-1}}\rangle\otimes|\psi_{x_{d}}\rangle$
of energy $E=\sum_{n=1}^{d}E_{x_{n}}$ at one of its zero-dimensional
corners. The resulting $d$-dimensional, $d$th-order topological
phase is characterized by a Cartesian-product invariant $\nu=(\nu_{1},\nu_{2},...,\nu_{d-1},\nu_{d})$,
with $\nu_{n}$ being the topological number classifying the phases of
1D subsystem $H_{x_{n}}$. It is clear that the corner state $|\psi\rangle$
could appear if and only if all components of $\nu$ are nontrivial,
so that every $|\psi_{x_{n}}\rangle$ exists as a topological edge
state of $H_{x_{n}}$, confirming the higher-order topology of the
system described by $H$.

The aforementioned scheme can be extended to realize non-Abelian HOTPs
by letting at least one element of the subsystems $\{H_{x_{n}}|n=1,2,...,d-1,d\}$
hold non-Abelian 1D topological states. The resulting
non-Abelian HOTPs can be further organized into two classes. The
first class, in which the subsystem $H_{x_{n}}$ is characterized
by a non-Abelian topological charge $Q_{x_{n}}$ for every $n=1,...,d$,
is referred to as pristine non-Abelian HOTPs. For example, let us
consider a system described by Hamiltonian $H=H_{x}\oplus H_{y}$,
with both $H_{x}$ and $H_{y}$ depicting 1D, three-band models of
non-Abelian topological insulators characterized by quaternion charges
$Q_{x}\in\mathbb{Q}_{8}$ and $Q_{y}\in\mathbb{Q}_{8}$ in the quaternion
group $\mathbb{Q}_{8}$. The 2D system $H$ could then support pristine
non-Abelian SOTPs, with each of them
characterized by a non-Abelian topological charge $\nu=(Q_{x},Q_{y})\in\mathbb{Q}_{8}\times\mathbb{Q}_{8}$
and having localized eigenmodes at the lattice corners if both $Q_{x}$
and $Q_{y}$ are different from the unit element of $\mathbb{Q}_{8}$.
The second class, in which a finite set of subsystems (e.g., $\{H_{x_{n}}|n=1,...,l;\,l<d\}$)
allow only Abelian topological phases, is referred to as hybridized
non-Abelian HOTPs. For example, let us still consider a system described
by $H=H_{x}\oplus H_{y}$, with $H_{x}$ ($H_{y}$) depicting
a 1D, three-band (two-band) model of non-Abelian (Abelian) topological
insulators characterized by quaternion charge $Q_{x}\in\mathbb{Q}_{8}$
(winding number $w_{y}\in\mathbb{Z}$). The 2D system $H$ would now
support hybridized non-Abelian SOTPs, with
each of them being described by a topological charge $\nu=(Q_{x},w_{y})\in\mathbb{Q}_{8}\times\mathbb{Z}$
and possessing corner-localized eigenstates if $Q_{x}$
is not the unit element and $w_{y}\neq0$. 
We interpret these boundary modes as corner states associated with NATC, and emphasize that the term ``non-Abelian'' here pertains to the eigenframe rotation of the multiband eigenvectors in momentum space, rather than the real-space braiding statistics typically associated with non-Abelian anyons. Therefore, the non-Abelian edge/corner states in our context refer to boundary states associated with bulk topological charges of non-Abelian characteristics. In the sense of bulk-boundary correspondence, a non-Abelian edge or corner state can only appear in our system if the non-Abelian component of bulk's pristine or hybridized NATC is nontrivial.

Before proceeding to explicit examples, let us analyze a
few more cases. Focusing on three dimensions, there
could be three types of non-Abelian third-order topological phases
with localized corner states and three types of non-Abelian SOTPs
with chiral hinge states according to our scheme.
The third-order non-Abelian topological phases, having
the common Hamiltonian $H=H_{x}\oplus H_{y}\oplus H_{z}$,
are pristine (hybridized) if all (one or two) of the subsystems
are characterized by non-Abelian (Abelian) topological charges. The
non-Abelian SOTPs have the common
Hamiltonian $H=H_{xy}\oplus H_{z}$, where $H_{xy}$ ($H_{z}$) represents
a 2D (1D) subsystem. When both $H_{xy}$ and $H_{z}$ are characterized
by NATCs, the system $H$ would support
pristine non-Abelian SOTPs with edge states along its hinges at the $xy$-surfaces
with $z=0$ and $z=L_{z}$ for a cubic lattice. Similar hinge states configurations
appear when $H_{xy}$ and $H_{z}$ describe non-Abelian
and Abelian topological insulators, respectively, with the key difference
that the resulting non-Abelian SOTPs are hybridized.
If $H_{xy}$ and $H_{z}$ separately describe
an Abelian Chern insulator and an NATI,
the system $H$ would support hybridized non-Abelian SOTPs
with chiral hinge modes lying along the boundaries
of their $xy$-surfaces with $z=0$ and $z=L_{z}$ for a cubic lattice.

In principle, this procedure can be carried forward to acquire $d$-dimensional,
$n$th-order non-Abelian HOTPs, either pristine or hybridized, with
$(d-n)$-dimensional boundary modes for any $1<n\leq d$. In the rest
of this study, we focus on the minimal case of non-Abelian HOTPs
characterized by the hybridized topological charge $\nu=(Q_{x},w)\in\mathbb{Q}_{8}\times\mathbb{Z}$
in two dimensions. Its simplest construction is illustrated in Fig.~\ref{fig: lattice model}.

According to the classification of HOTPs~\cite{HOTI5}, the model constructed by our approach belongs to an ``extrinsic'' class. Although the 1D subsystems exhibit bulk-gap-protected topology characterized by winding numbers or NATCs, the resulting zero-dimensional corner states are primarily protected by the lower-dimensional boundary gaps generated by the coupled-wire structure rather than by additional crystalline symmetries. The occurrence of a phase transition between different HOTPs then requires the closure of both the bulk and boundary gaps. This extrinsic nature provides a versatile platform for quantum simulations. By independently tuning the coupling parameters, one can control the topological phase transitions within each subsystem. Notably, in our hybridized non-Abelian HOTPs, this framework allows the flexible combination of Abelian and non-Abelian components, enabling programmable control of distinct topological phase transitions as well as selective manipulation of higher-order boundary modes.

In the coupled-wire construction, the full energy spectrum of a 2D system is given by $E = E_x + E_y$. While our scheme ensures the emergence of corner modes from the topological edge states of the 1D subsystems, it may lead to regimes where the corner modes are energetically immersed in the bulk or edge spectral continuum.	To accurately describe this, we refer to these regimes as SOTPs, where the stability of corner modes is rooted in the subsystem symmetries rather than a global spectral gap.

We note that non-Abelian HOTPs may also be realized by exploiting the geometric nature of NATC.
For a 1D system, the Hamiltonian in momentum space can be written as $H(k_{x})=R(k_{x})\Lambda R^{\top}(k_{x})$, where $\Lambda$ is the eigenvalue matrix and $R(k_{x})$ contains eigenvectors. A 2D version can be realized with the Hamiltonian $H(k_{x},k_{y}) = R_{2}(k_{y}) R_{1}(k_{x}) \Lambda R^{\top}_{1}(k_{x})R^{\top}_{2}(k_{y})$, which extends the non-Abelian structures of eigenvectors in $R_1(k_{x})$ and $R_2(k_{y})$ to two dimensions~\cite{HONATI1}. This construction allows the non-Abelian topology along two momentum directions $k_x$ and $k_y$ to be characterized independently. One may associate a NATC $Q_{x}$ with the eigenframe rotation of $R_1(k_{x})$ versus $k_{x}$ and another NATC $Q_{y}$ with the rotation of $R_2(k_{y})$ along $k_{y}$. The global NATC is given by $q = (Q_{x}, Q_{y})$, which cannot appear in arbitrary combinations in a gapped system. To see this, let us initialize the 2D state at $(k_{x},k_{y})=(-\pi,-\pi)$. If this state is first transported along $k_{x}$ and then along $k_{y}$ across the Brillouin zone to reach $(\pi,\pi)$, the resulting rotation matrix is $Q_{y}Q_{x}$. Instead, if the evolution is performed first along $k_y$ and then $k_x$, the resulting matrix reads $Q_{x}Q_{y}$. These two paths, although connecting the same initial and final points, lead to different results if $[Q_{x},Q_{y}]\neq0$. This could only happen when the energy bands exhibit degeneracies. If the chosen path in $k$-space crosses such a degeneracy, the NATC may change abruptly, allowing noncommuting charges to appear simultaneously. 
Following this scheme, corner states depicted by NATCs can emerge in 2D four-band systems characterized by two independent group elements in the generalized quaternion group ${\mathbb Q}_{16}$. For systems described by the quaternion group ${\mathbb Q}_{8}$, only two distinct types of edge state can appear~\cite{HONATI1}.

\subsection{Model}\label{sec: Model}
We now propose a minimal model of non-Abelian HOTPs by stacking 1D trimer chains along a second dimension, where the interchain coupling strengths alternate along the stacking direction, as illustrated in Fig.~\ref{fig: lattice model}(a). The Hamiltonian of the resulting 2D lattice model reads
\begin{equation}
    \begin{split} H&=\sum_{n,m=1}^{N,M/2}\sum_{\alpha,\beta=A,B,C}\big[ (J_{1}\ket{n,2m,\alpha}\bra{n,2m-1,\alpha}\\
    &+J_{2}\ket{n,2m+1,\alpha}\bra{n,2m,\alpha} +\text{h.c.})\\   
    &+S_{\alpha}\ket{n,m,\alpha}\bra{n,m,\alpha}\\
    &+(J_{\alpha\beta}\ket{n+1,m,\beta}\bra{n,m,\alpha}+\text{h.c.})\big ],
    \end{split}\label{eq: Full Hamiltonian}
\end{equation}
where $n$ and $m$ denote unit cell and lattice indices along $x$ and $y$ directions. $N$ ($M/2$) denotes the number of unit cells along $x$ ($y$). $\alpha$ and $\beta$ label sublattice degrees of freedom along $x$. The terms $J_{\alpha \beta}$ represent nearest-neighbor (NN) hopping amplitudes along $x$, with the constraint that $J_{AC}=J_{CA}=0$. Explicitly, we set $J_{AB} = J_{BA} = \mathrm{i}u$ and $J_{BC} = J_{CB} = \mathrm{i}v$. $S_{\alpha}$ denotes the strength of onsite potential in sublattice $\alpha$. $J_1$ and $J_2$ are NN hopping amplitudes along $y$-direction. The Hilbert space of the 2D system is $3N\times M$-dimensional.
The Hamiltonian in Eq.~(\ref{eq: Full Hamiltonian}) has the Kronecker sum structure of Eq.~(\ref{eq:GeneralH}), which allows us to express it as
\begin{equation}
    H=H_{x}\oplus H_{y}=H_{x}\otimes {\mathbb I}_{y}+{\mathbb I}_{x}\otimes H_{y}.\label{eq: Hamiltonian Kron}
\end{equation}
Here, $H_x$ and $H_y$ represent 1D subsystems along $x$ and $y$ directions. They are explicitly given by
\begin{align}
    \label{eq: x Hamiltonian OBC}
    H_{x}&=\sum_{n=1}^{N}\sum_{\alpha,\beta=A,B,C}\big[S_{\alpha}\ket{n,\alpha}\bra{n,\alpha}\nonumber\\
        &+(J_{\alpha\beta}\ket{n+1,\beta}\bra{n,\alpha}+\text{h.c.})\big],\\
    H_{y}&=\sum_{m=1}^{M/2}(J_{1}\ket{2m}\bra{2m-1}\nonumber\\
    &+J_{2}\ket{2m+1}\bra{2m}+\text{h.c.}).\label{eq: y Hamiltonian OBC}
\end{align}
The eigenvalue equation of 2D Hamiltonian $H$ is given by $H\ket{\Psi_{q}}=E_{q}\ket{\Psi_{q}}$, where $E_{q}$ is the energy of its $q$th eigenstate $\ket{\Psi_{q}}$.
For the decoupled 1D Hamiltonian $H_{x(y)}$, the eigenvalue equation reads $H_{x(y)}\ket{\Psi_{x(y)}}=E_{x(y)}\ket{\Psi_{x(y)}}$.
The spectrum of each 1D Hamiltonian is illustrated in Sec.~\ref{sec: 1D Hamiltonian}.
The Kronecker sum structure in Eq.~(\ref{eq: Hamiltonian Kron}) implies that for any eigenstate $|\Psi_q\rangle$ of $H$
with energy $E_q$, there must be eigenstates $|\Psi_x\rangle$ and $|\Psi_y\rangle$ of the subsystems $H_x$ and $H_y$ with
energies $E_x$ and $E_y$, such that $|\Psi_q\rangle=|\Psi_x\rangle\otimes|\Psi_y\rangle$ and $E_q=E_x+E_y$.
One could then deduce the eigensystem of $H$ from the eigenstates and eigenenergies
of its 1D subsystems $H_x$ and $H_y$. This further allows our 2D system to inherit the topology of its lower dimensional
components via hybridizing their non-Abelian and/or Abelian topological characters.

Under OBC, if $H_{x}$ has an edge state $\ket{\Psi_{xe}}$ with energy $E_{xe}$, and if $H_{y}$ has an edge state $\ket{\Psi_{ye}}$ with energy $E_{ye}$, the 2D Hamiltonian $H$ in Eq.~(\ref{eq: Full Hamiltonian}) would necessarily host a corner-localized state $\ket{\Psi_c}=\ket{\Psi_{xe}} \otimes \ket{\Psi_{ye}}$ with energy $E_{c}=E_{xe}+E_{ye}$.
In our case, the Hamiltonian $H_{y}$ along $y$-direction is just the well-known SSH model, whose topological phase hosts two degenerate edge zero modes protected by chiral symmetry. Therefore, we have $E_{ye}=0$ and the energy of corner states is $E_{c}=E_{xe}$, relying only on the edge state of $H_{x}$.
Moreover, the number and degeneracy of corner states are primarily determined by the edge spectrum of $H_{x}$, as the SSH Hamiltonian $H_{y}$ contributes only a fixed two-fold degeneracy for each corner state. For example:

1) If $H_{x}$ has one pair of degenerate edge states, the 2D system $H$ will exhibit four degenerate corner states.

2) If $H_{x}$ has two pairs of edge states at different energies, the 2D system $H$ will have two sets of fourfold degenerate corner states.

3) If $H_{x}$ has four edge modes at the same energy, the corner states of $H$ will be eightfold degenerate.

More detailed evidence and numerical results are presented in the next two sections.

\section{Topological invariants and phase diagram}\label{sec: topological invariant}
This section first introduces the topological characterization for each of our 1D subsystems
that make up the 2D minimal lattice model $H$. We obtain the topological invariants, edge modes, and
bulk-edge correspondence for both subsystems $H_x$ and $H_y$. Building on these, we define the hybridized 
topological invariants for our 2D model and obtain its phase diagram, which shows a rich set of 
non-Abelian SOTPs.

\subsection{Topological invariants of 1D subsystems}
The direct sum structure of Hamiltonian $H$ permits a natural definition of its topological invariants, thereby offering a link between the first and second order topology. Under periodic boundary condition (PBC), the Bloch Hamiltonians of $H_x$ and $H_y$ are given by
\begin{alignat}{1}
H_x(k_{x})& = \begin{pmatrix}V_{AA} & 2u\sin k_{x} & 0\\
2u\sin k_{x} & V_{BB} & 2v\sin k_{x}\\
0 & 2v\sin k_{x} & V_{CC}
\end{pmatrix},\label{eq:Hkx}\\
H_y(k_{y})& = (J_{1}+J_{2}\cos k_{y})\sigma_{1}+J_{2}\sin k_{y}\sigma_{2}\nonumber\\
& = d_{1}(k_{y})\sigma_{1}+d_{2}(k_{y})\sigma_{2}.\label{eq:Hky}
\end{alignat}
Here, $V_{\alpha\alpha}\equiv S_\alpha+2J_{\alpha\alpha}\cos{k_x}$ for $\alpha=A,B,C$. $\sigma_{1,2,3}$ are the $x,y,z$ components of Pauli matrices. Along the $y$ direction, we have $H_y(k_{y})\ket{\psi_{\pm}(k_{y})}=E_{\pm}\ket{\psi_{\pm}(k_{y})}$, where $E_{\pm}=\pm E_{y}$ is due to chiral symmetry. Along the $x$ direction, we have $H_x(k_{x})\ket{\psi_{p}(k_{x})}=E_{p}\ket{\psi_{p}(k_{x})}$, where $p=1,2,3$. The states $\ket{\psi_{p}(k_{x})}$ and $\ket{\psi_{\pm}(k_{y})}$ are defined in two Hilbert subspaces $\mathcal{H}_{k_{x}}$ and $\mathcal{H}_{k_{y}}$, respectively.
They have distinct geometric and topological structures within their respective parameter manifolds $\mathcal{M}_x$ and $\mathcal{M}_y$.
The fundamental homotopy group of them can be expressed as $\pi_{1}(\mathcal{M}_{x})=\mathbb{Q}_{8}$~\cite{NATI1} and $\pi_{1}(\mathcal{M}_{y})=\mathbb{Z}$. The 2D Bloch Hamiltonian reads
\begin{equation}
    H(\mathbf{k})=H_x(k_{x})\otimes\sigma_{0}+{\mathbb I}_{3}\otimes H_y(k_{y}),
\end{equation}
where $\sigma_{0}$ and ${\mathbb I}_{3}$ denote $2\times2$ and $3\times3$ identity matrices. As a result, the 2D model has six bulk bands, whose eigenenergies are $\pm E_{py}=E_{p}\pm E_{y}$ for $p=1,2,3$ and the related eigenstates $\ket{\Psi_{p\pm}(\mathbf{k})}=\ket{\psi_{p}(k_{x})} \otimes \ket{\psi_{\pm}(k_{y})}$.
The corresponding parameter space forms the Cartesian product $\mathcal{M} = \mathcal{M}_x \times \mathcal{M}_y$, with each point ${\bf k} \in \mathcal{M}$
associated with a quantum state $\ket{\Psi_{p\pm}(\mathbf{k})}$.
This construction naturally defines a composite manifold structure in which the local state is induced by the tensor product of two fiber bundles, providing a geometric framework for describing coupled and multi-layer topological systems.
In Fig.~\ref{fig: lattice model}(b), we present the 2D Brillouin zone (BZ) of $H({\bf k})$ on a torus with $k_{x},k_{y}\in[-\pi,\pi]$.
Each point $\mathbf{k}$ on the surface spans six eigenenergies and eigenstates.
Taking advantage of the tensor-product structure of $\ket{\Psi_{p\pm}({\mathbf{k}})}$, we can study the contribution of the two subsystem Hamiltonians $H_x(k_{x})$ and $H_y(k_{y})$ to the topological properties of the 2D lattice model separately.

\begin{figure}
    \centering
    \includegraphics[width=0.9\linewidth]{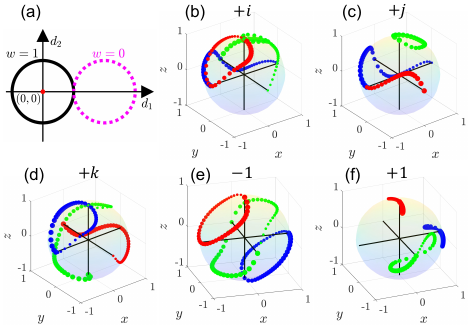}
    \caption{Geometric illustrations of the Abelian winding number and non-Abelian quaternion charge. (a) shows the trajectory of Bloch vector $\mathbf{d}(k_{y})$ as $k_{y}$ varies from $-\pi$ to $\pi$. In the topological (trivial) phase with winding number $w=1$ ($w=0$), the trajectory of $\mathbf{d}(k_{y})$ in black solid line (pink dashed line) encircles once (does not encircle) the origin. (b)-(f) show the eigenframe rotation associated with each quaternion charge. The eigenstates $\ket{\psi_{1,2,3}(k_{x})}$ are marked by red, green, and blue dots at every $k_{x}$. The size of dots in each group increases progressively from $k_{x}=-\pi$ to $\pi$.}
    \label{fig: Topological geometry}
\end{figure}

Along the meridional direction of the tours, we connect the ends of BZ along the $y$ direction, such that the quasimomentum $k_{y}$ forms a closed loop for each $k_{x}$, as illustrated by the purple loop in Fig.~\ref{fig: lattice model}(b). The eigenstates $\{\ket{\psi_p(k_{x})}\}$ are independent of $k_{y}$, so that only $\ket{\psi_\pm(k_{y})}$ contributes to the topological invariant along this loop. According to the construction of our 2D system, the Hamiltonian $H_{y}$ takes the form of an SSH model, which belongs to the BDI symmetry class \cite{REVTI3}. The topological properties of $H_{y}$ can be characterized by a winding number due to its chiral symmetry $\Gamma=\sigma_3$, i.e.,
\begin{equation}\label{eq: winding number}
    w=\frac{1}{2\pi} \int^\pi_{-\pi} \frac{d_{1}(k_{y})\partial_{k_{y}} d_{2}(k_{y})-d_{2}(k_{y})\partial_{k_{y}} d_{1}(k_{y})}{|\mathbf{d}(k_{y})|^{2}} dk_{y},
\end{equation}
where $\mathbf{d}(k_{y})=(d_1(k_y),d_2(k_y))$. The $w$ has a geometric interpretation: it counts the number of times the Bloch vector $\mathbf{d}(k_{y})$ winds around the origin of $d_1$-$d_2$ plane as $k_{y}$ varies from $-\pi$ to $\pi$, as illustrated in Fig.~\ref{fig: Topological geometry}(a). There are two distinct phases for $H_y(k_{y})$. In the topological insulator phase with $w=1$ (black line), the Hamiltonian $H_{y}$ hosts two zero-energy edge states under OBC [Fig.~\ref{fig: 1D spectrum}(a)]. In the trivial insulator phase with $w=0$ (magenta dashed line), no edge states appear. At phase transition ($J_1=J_2$), the origin lies on the trajectory of $\mathbf{d}(k_{y})$, rendering the winding number ill-defined.

The $w$ in Eq.~(\ref{eq: winding number}) is mathematically equivalent to the Zak phase $\gamma_{y}$. The latter can also serve as a topological invariant~\cite{asbothTI}, which is equal to $0$ ($\pi$) in the topologically trivial (nontrivial) phase of the SSH model. However, for the multi-band Hamiltonian $H_x(k_{x})$ along $x$, the Zak phases $\gamma_{x}=(\gamma_{12},\gamma_{23})$ of the two band gaps are not enough to characterize its topological phases \cite{NATI1}. For example, $\gamma_{x}=(0,\pi)$ indicates a pair of degenerate edge states between the second and third bands, $\gamma_{x}=(\pi,0)$ leads to a pair of degenerate edge states between the first and second bands, $\gamma_{x}=(\pi,\pi)$ implies the existence of two pairs of edge states in both band gaps, and $\gamma_{x}=(0,0)$ signifies the absence of edge states. Nevertheless, this characterization remains incomplete for $H_x(k_{x})$, as edge states can still emerge even when $\gamma_{12}=\gamma_{23}=0$ (see details in Appendix \ref{app: Zak phase} and Sec.~\ref{sec: 1D Hamiltonian}). 
NATCs are then required to capture these additional topological features, offering a complete description of the multigap topological phases \cite{NATS1,NATI1}.

The quasimomentum $k_{x}$ is set along the azimuthal direction of the torus, forming a BZ as illustrated by the orange loop in Fig.~\ref{fig: lattice model}(b).
For each $k_{x}$ loop, the eigenstates $\{\ket{\psi_\pm(k_y)}\}$ remain fixed, so the topological properties are fully determined by the Hamiltonian $H_x(k_{x})$ along $x$ direction.
Here, conventional Abelian topological invariants, such as the geometric phase, are insufficient to characterize all distinct phases of $H(k_x)$.
In this case, a matrix-valued noncommutative NATC can be introduced \cite{NATS1}.
For each eigenstate of our $\mathcal{{PT}}$-symmetric Hamiltonian $H_x(k_{x})$, its three components $\ket{\psi_{p}(k_{x})}=[\psi_{p1}(k_{x}), \psi_{p2}(k_{x}), \psi_{p3}(k_{x}) ]^{\top}$ are real numbers, forming a three-dimensional (3D) eigenframe $(x,y,z)$.
With the change of $k_{x}$, the eigenframe rotates in 3D Euclidean space~\cite{NATI1}. The geometric features of the resulting NATCs are illustrated in Figs.~\ref{fig: Topological geometry}(b)--\ref{fig: Topological geometry}(f), where the eigenstates $\ket{\psi_{1,2,3}(k_{x})}$ are marked by red, green and blue dots for each $k_{x}$.
As $k_x$ goes from $-\pi$ to $\pi$, the size of the dots gradually increases. There are five distinct cases in total.
(i) Two of the three vectors rotate by $\pi$ around the third [three distinct configurations in Figs.~\ref{fig: Topological geometry}(b)--\ref{fig: Topological geometry}(d)], and the counter-rotations constitute their conjugate class; (ii) two of them rotate by $2\pi$ around the third [Fig.~\ref{fig: Topological geometry}(e)]; and (iii) there are no rotations around any vectors [Fig.~\ref{fig: Topological geometry}(f)]. For the latter two cases, the initial and final eigenframes are identical after rotation. As a result, they cannot be distinguished by conventional geometric phases. However, they exhibit distinct properties: the former corresponds to a topological phase, while the latter represents a trivial one.

Under the given symmetry constraints, the eight geometric configurations above yield all possible topological structures of $H_x(k_{x})$ with distinct NATCs, which are defined by the generalized Wilson loop
\begin{equation}
    W=\hat{\mathsf{P}}\exp\left[ \oint _{-\pi}^{\pi} A(k_{x}) dk_{x} \right],
\end{equation} 
where $\hat{\mathsf{P}}$ is the path-ordering operator. The affine Berry-Wilczek-Zee (BWZ) connection $A(k_{x})$ is given by
\begin{equation}
    [A(k_{x})]_{q,q'}=\braket{\psi_{q}(k_{x}) | \partial_{k_{x}}|\psi_{q'}(k_{x})},
\end{equation}
which is antisymmetric and can be decomposed by SO(3) generators, i.e., $A(k_{x})=\sum_{p=1}^{3}\beta_p L_p$ for $p=1,2,3$.
Using SU(2) as a double cover of SO(3), we can replace $L_{p}$ by $t_{p}=-\frac{\mathrm{i}}{2}\sigma_{p}$, yielding the affine BWZ connection $\bar{A}(k_{x})=\sum_{p=1}^{3}\beta_{p}t_{p}$.
The NATC $q_{x}$ accumulated along the BZ of $k_x$ is then given by
\begin{equation}\label{eq: NATC}
    q_{x}=\hat{\mathsf{P}}\exp\left[ \oint _{-\pi}^{\pi} \bar{A}(k_{x}) dk_{x} \right].
\end{equation}
The geometric structure of non-Abelian topological states has eight distinct classes, allowing the NATCs $q_{x}$ to have eight matrix values from different rotations around the axes, i.e.,
\begin{equation}
    q_{x}=\{ \sigma_{0}, \pm \mathrm{i}\sigma{_{x}},\pm \mathrm{i} \sigma_{y}, \pm \mathrm{i} \sigma_{z}, -\sigma_{0}\}.
\end{equation}
It is isomorphic to the quaternion group, given by
\begin{equation}
    Q_{x}= \{ 1,\pm i, \pm j, \pm k, -1 \}.
\end{equation}
The $\pm\pi$ rotations around the $x/y/z$ axes correspond to the charges $\pm i/\pm j/\pm k$. The $\pm 2\pi$ rotations around the $x/y/z$ axes correspond to the charge $(\pm i)^{2}=(\pm j)^{2}=(\pm k)^{2}=-1$.
The topological phases of $H_x(k_{x})$ are therefore depicted by these NATCs. The elements $\pm i$, $\pm j$ and $\pm k$ belong to different conjugacy classes. The charge $-1$ corresponds to a topological phase that cannot be described by conventional geometric phases. In contrast, the trivial phase is characterized by the identity element $1$.
The NATCs $i$, $j$, $k$ and $1$ map to the Zak phases $(0,\pi)$, $(\pi,\pi)$, $(\pi,0)$ and $(0,0)$, respectively. The NATC $-1$ cannot be described by Zak phases, highlighting the necessity of non-Abelian characterizations~\cite{NATS1,NATI1}.

\subsection{Edge states of 1D subsystems}\label{sec: 1D Hamiltonian}

\begin{figure}
    \centering   
    \includegraphics[width=0.9\linewidth]{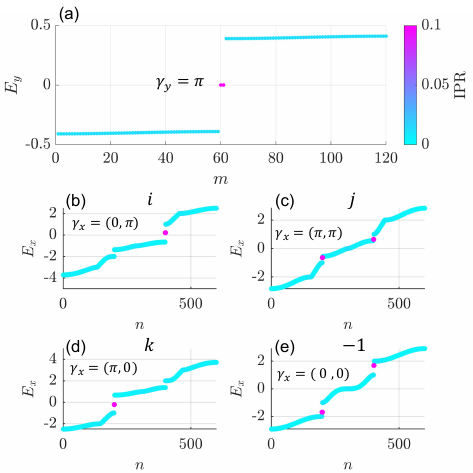}
    \caption{Spectrum of 1D subsystems $H_x$ and $H_y$. All panels share the same color bar, which indicates the magnitude of IPR for each state. (a) shows the spectrum of $H_{y}$ in its topologically nontrivial region with $J_{1}=0.01$ and $J_{2}=0.4$. (b)--(e) show the spectrum and IPR of $H_{x}$. We set $S_{C}=2\cos{\varphi}$ and $J_{CC}=\frac{1}{2}+\frac{1}{2}\sin{\varphi}$ with $\varphi=\pi$ for (b), $\varphi=3\pi/2$ for (c), $\varphi=0$ for (d), and $\varphi=\pi/2$ for (e). Other parameters are $J_{AA}=-1$, $2J_{BB}=-u=v=1$, and $S_{A}=S_{B}=0$.}
    \label{fig: 1D spectrum}
\end{figure}

The Kronecker sum structure of 2D Hamiltonian $H$ allows its eigensystem to be expressed as the sum and tensor product of eigenvalues and eigenstates of 1D Hamiltonians $H_{x}$ and $H_{y}$ along two directions ($E = E_x + E_y, \ket{\Psi} = \ket{\Psi_x} \otimes \ket{\Psi_y}$).
Consequently, the topological properties of the 2D system can be deduced from those of the 1D subsystems.
Here, we present their topological characteristics in order to clarify the non-Abelian origin of corner and edge states in our 2D minimal model. 
The explicit forms of $H_{y}$ and $H_{x}$ are given by Eqs.~(\ref{eq: x Hamiltonian OBC}) and (\ref{eq: y Hamiltonian OBC}). 
Their corresponding energy spectra are shown in Fig.~\ref{fig: 1D spectrum}(a) and Figs.~\ref{fig: 1D spectrum}(b)--\ref{fig: 1D spectrum}(e) under OBC.

The Hamiltonian $H_y(k_y)$ in Eq.~(\ref{eq: x Hamiltonian OBC}) describes an SSH model. When $|J_{1}|<|J_{2}|$ the system enters a nontrivial topological phase. It hosts a pair of degenerate edge modes at zero energy under OBC, characterized by a winding number $w=1$ or a Zak phase $\gamma=\pi$ under PBC. When $|J_{1}|>|J_{2}|$ the system becomes topologically trivial and exhibits no edge states, corresponding to $w=0$ or $\gamma=0$. The topological phase transition occurs at $|J_{2}|=|J_{1}|$.
The spectrum and inverse participation ratio (IPR) of $H_{y}$ can be obtained from the eigenvalue equation $H_{y}\ket{\Psi_{y}}=E_{y}\ket{\Psi_{y}}$ and ${\rm IPR}=\sum_{m=1}^{L_y}|\langle m|\Psi_{y}\rangle|^{4}$. The resulting spectrum in the topologically nontrivial regime is illustrated in Fig.~\ref{fig: 1D spectrum}(a). It includes a pair of edge states (in pink) at $E_{y}=0$, which are protected by the chiral symmetry of $H_{y}$. All others are bulk states, whose IPR values tend to zero in the limit $L_y\rightarrow\infty$.

The Hamiltonian $H_{x}$ along $x$-direction is given by Eq.~(\ref{eq: x Hamiltonian OBC}), which satisfies $H_{x}\ket{\Psi_{x}}=E_{x}\ket{\Psi_{x}}$.
We present the spectrum of $H_{x}$ with NATCs $Q_{x}=i,j,k,-1$ under OBC in Figs.~\ref{fig: 1D spectrum}(b)--\ref{fig: 1D spectrum}(e).
The pink dots highlight edge states with large IPR.
In Figs.~\ref{fig: 1D spectrum}(b) and \ref{fig: 1D spectrum}(d), the upper and lower band gaps each hold two degenerate edge states. These cases correspond to the NATCs $i$ and $k$. The Zak-phase approach remains valid for them, with the corresponding values $\gamma_{x}=(0,\pi)$ and $(\pi,0)$.
Two pairs of doubly degenerate edge states are observed in Fig.~\ref{fig: 1D spectrum}(c). Each of them is located in a different band gap, corresponding to the case with NATC $j$. Here, the Zak phase could also capture the existence of these edge states.
In Fig.~\ref{fig: 1D spectrum}(e), although two pairs of doubly degenerate edge states emerge in the two gaps, they correspond to the NATC $-1$.
In this case, the Zak-phase approach breakdowns, as both band gaps exhibit trivial Zak phases $\gamma_{x}=(0,0)$ despite the presence of topological edge states.
It demonstrates that the Zak phase is insufficient to classify non-Abelian topological phases and highlights the necessity of NATCs.
Further details of the anomalous edge states of phase $-1$ are shown in Appendix \ref{app: another -1}. They can be regarded as phases generated by $i^{2}$, $j^{2}$, and $k^{2}$, where the corresponding edge states appear in their respective band gaps, leading to the doubling of edge state numbers in Fig.~\ref{fig: other -1}.

In essence, the corner states of 2D system $H$ originate from the hybridization of Abelian topology (the winding number of $H_{y}$) and non-Abelian topology (the NATC of $H_{x}$). This establishes our minimal model as a realization of non-Abelian SOTPs, whose phase structure will be revealed in the next subsection.

\subsection{Hybridized invariant for higher-order topology}
The above discussions have identified suitable topological invariants for our subsystem Hamiltonians.
A system formed by their direct sum should carry the product topology of the quaternion charge and winding number of $H_x(k_x)$ and $H_y(k_y)$.
The resulting topological invariant, as elements of the SU(2)$\times$U(1) group, takes the form
\begin{equation}
    \nu = (Q_x,w)\in{\mathbb Q}_8\times{\mathbb Z},
\end{equation}
where $Q_x$ and $w$ denote the NATC and winding number of $H_x(k_x)$ and $H_y(k_y)$.
The multiplication rule between two such hybridized NATCs, as acquired by a direct product group $G=\mathbb{Q}_{8}\times\mathbb{Z}$, is given by
\begin{equation}
    \nu_{1}\circ\nu_{2}\equiv(Q_{x1}Q_{x2},w_{1}+w_{2}).
\end{equation}
The winding number $w\in{\mathbb Z}$ is additive and obeys the commutative law $w_{1}+w_{2}=w_{2}+w_{1}$. The additivity of Abelian topological invariants implies that the total topological charge can typically be obtained by summing the contributions from subsystems, or, in multiband systems, by accumulating the topological invariants of all occupied bands.
Meanwhile, the quaternions are not commutative, e.g., $ij=-ji=k$.
Therefore, the topological charge $\nu$ also does not satisfy the commutative law for any $Q_{x1},Q_{x2}\in \{\pm i,\pm j,\pm k\}$.
All possible values of $\nu$ thus form a non-Abelian group.
In summary, the SOTPs in our 2D system $H$ can be fully characterized by a hybridized NATC $\nu$.
Accordingly, four distinct configurations of topological charge could arise depending on the values of the two components of $\nu$:  
(i) both $w$ and $Q_x$ are nontrivial,  
(ii) only the quaternion charge $Q_{x}$ is nontrivial,  
(iii) only the winding number $w$ is nontrivial, and  
(iv) both invariants are trivial.

\begin{figure}
    \centering
    \includegraphics[width=0.9\linewidth]{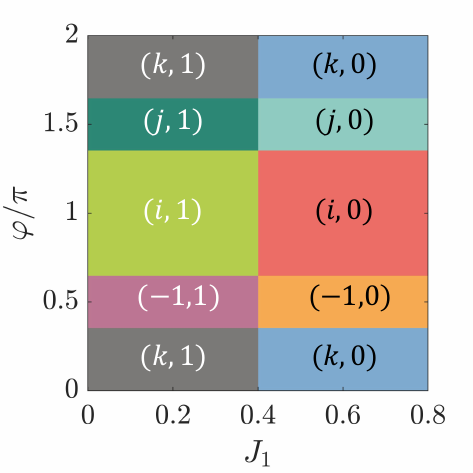}
\caption{Phase diagram of 2D non-Abelian HOTPs characterized by hybridized topological charge $\nu=(Q_x,w)$. Different SOTPs are indicated by regions with distinct colors, and their topological invariants shown therein. Other system parameters are the same as those in Fig.~\ref{fig: 1D spectrum}.}\label{fig: Phase_Diagram}
\end{figure}

For case (i), since both topological charges are nontrivial, multiple distinguishable topological phases occur due to the quaternion charge $Q_x$. These phases are characterized by hybridized invariant $\nu = (\pm i,1)$, $(\pm j,1)$, $(\pm k,1)$ and $(-1, 1)$.
A phase diagram containing all these phases is shown in Fig.~\ref{fig: Phase_Diagram}.
In case (ii), while the winding number $w$ is trivial, the system hosts a family of first-order topological phases characterized by nontrivial $Q_x$. They are weak topological phases with non-Abelian edge bands, which can also be identified in Fig.~\ref{fig: Phase_Diagram}.
For case (iii), the quaternion charge becomes trivial while the winding number $w$ is nontrivial.
The topology is then governed by the 1D winding number, showing only Abelian edge bands.
Cases (ii) and (iii) form a complementary pair, each capturing only one aspect of the hybridized invariant $\nu = (Q_x,w)$.
Finally, both the winding number $w$ and quaternion charge $Q_x$ are trivial in case (iv), making the hybridized invariant $\nu=(1,0)$.
This regime represents a trivial phase without any signatures of first- and second-order topology.

The emergence of corner and edge states under OBC depends on the interplay of invariants $Q_x$ and $w$. Hybridized corner states appear exclusively in case (i), where both the $w$ and $Q_x$ are nontrivial. Case (ii) hosts weak non-Abelian edge bands when OBC and PBC are taken along the $x$ and $y$ directions, while case (iii) exhibits weak Abelian edge bands when OBC and PBC are taken along the $y$ and $x$ directions. Case (iv) does not support boundary states. A detailed analysis of these edge and corner states is given in Sec.~\ref{sec: spectrum OBC}.

The hybridized topological invariant $\nu$ is calculated using Eqs.~(\ref{eq: winding number}) and (\ref{eq: NATC}). 
In the resulting phase diagram Fig.~\ref{fig: Phase_Diagram}, when $J_{1}<0.4$,
the winding number $w=1$ and the quaternion charge $Q_{x}$ take nontrivial values.
In this regime, both $H_x(k_{x})$ and $H_y(k_{y})$ are topologically nontrivial, endowing the full 2D model with a family of distinguishable non-Abelian HOTPs.
These regimes host both gapped and gapless topological phases. In the later case, the bulk bands are only separated by direct gaps but may hold no indirect gaps in momentum space.
These phases differ by the geometry of eigenstates $\ket{\Psi_{p\pm}({\bf k})}$, where $\ket{\psi_{\pm}(k_{y})}$ share the same winding number $w$ but $\ket{\psi_{p}(k_{x})}$ undergoes distinct rotations for different values of $\varphi$.
As discussed in Sec.~\ref{sec: corner states}, these differences are directly manifested in the energy distribution of corner states under OBC.
When $J_{1}>0.4$, the winding number $w=0$, and $H_y(k_{y})$ becomes topological trivial.
In this region, only weak non-Abelian topological edge bands could survive under OBC (see Sec.~\ref{sec: y trivial} for details).

Non-Abelian SOTPs with charge $\nu=(-1,1)$ can be realized through multiple mechanisms, and a representative case is presented in Fig.~\ref{fig: Phase_Diagram}. In particular, different $\nu=(-1,1)$ phases can be transformed into each other without going through band touchings, indicating that they are topologically equivalent. The distinctions among their corner states are discussed in Appendix \ref{app: another -1}. These SOTPs characterized by $\nu=(-1,1)$ further demonstrate the necessity of employing NATC for their characterization. For the 1D subsystem along $x$ direction, any of its phases with $Q_x=-1$ cannot be described by the conventional geometric phase, as shown in Fig.~\ref{fig: 1D spectrum}(e) and Figs.~\ref{fig: other -1}(a1)--\ref{fig: other -1}(c1). Since the global topology of our 2D system is determined jointly by the topological invariants of two 1D subsystems, the NATC must be included in the hybridized topological invariant, which consequently gives rise to its noncommutative nature.

To summarize, we have introduced a hybridized invariant $\nu=(Q_x,w)$, constructed from both Abelian and non-Abelian topological charges, to characterize 2D SOTPs of non-Abelian origins. Moreover, our minimal model supports both Abelian (with the $Q_x$ component unchanged) and non-Abelian (with the $w$ component unchanged) topological transitions via changing system parameters, bridging two seemingly incompatible topological classes into a unified framework. This unification not only enriches the landscape of HOTPs, but also yields more exotic bulk-boundary correspondences, as will be discussed in the next section.

\section{Bulk-edge-corner correspondence}\label{sec: spectrum OBC}
In Sec.~\ref{sec: Model}, we performed a qualitative analysis on the topological edge and corner states of our 2D lattice model.
In this section, we proceed to a more thorough discussion of these boundary modes and their relations to the hybridized non-Abelian topological invariants of the bulk. We first consider in Sec.~\ref{sec: corner states} the corner states in cases where both $H_{x}$ and $H_{y}$ are set in topologically nontrivial regimes.
Subsequently, we reveal weak topological edge bands arising from the coupling between topological edge states in one dimension and trivial bulk states in the other in Sec.~\ref{sec: y trivial}.

\subsection{Non-Abelian topological corner states}\label{sec: corner states}

\begin{figure*}
    \centering
    \includegraphics[width=0.95\linewidth]{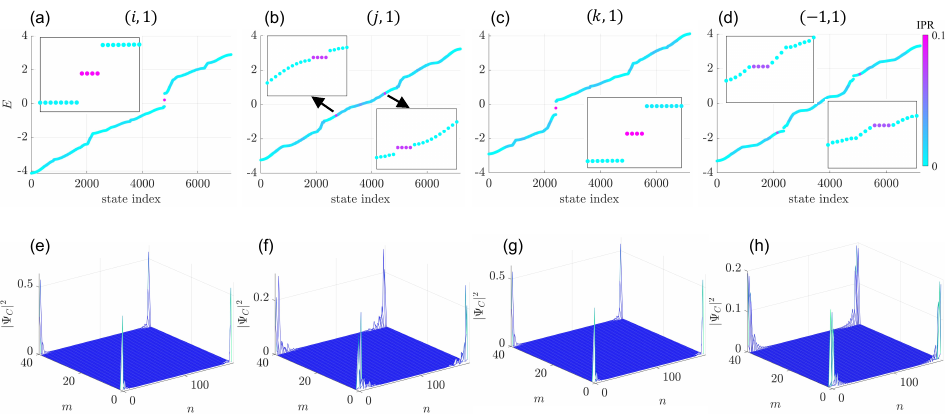}
    \caption{Spectrum and corner states of the minimal hybridized non-Abelian model under OBC along both dimensions. (a)--(d) show the spectra of the system for different values of $\varphi$. The insets highlight the energy windows of corner states. The shared color bar indicates the IPR of each eigenstate. (e)--(h) show the probability distributions of corner states in the spectra (a)--(d). We set $J_{1}=0.01$, $J_{2}=0.4$, and $\varphi=\pi$ for (a), (e); $\varphi=3\pi/2$ for (b), (f); $\varphi=0$ for (c), (g); and $\varphi=\pi/2$ for (d), (h). Other parameters are equal to those of Fig.~\ref{fig: Phase_Diagram}. The lattice size is $N=60$ and $M=40$.}\label{fig: band and corner}
\end{figure*}

The theory developed in the last section implies
that corner states associated with NATC could appear in our 2D system described
by $H$ if and only if the quaternion charge and winding number of
its 1D subsystems satisfy $Q_{x}\neq1$ and $w\neq0$. With the knowledge
of edge states of $H_{x}$ and $H_{y}$ in mind, we could arrive at
the following rules of bulk-corner correspondence for our minimal
model, i.e.,
\begin{equation}
N_{c}=\begin{cases}
4|w||Q_{x}^{2}|, & Q_{x}=\pm i,\pm k,\\
8|w||Q_{x}^{2}|, & Q_{x}=\pm j,-1,\\
0, & Q_{x}=1,
\end{cases}\label{eq:BBCmin}
\end{equation}
where $N_{c}$ denotes the total number of corner states for each
given set of bulk topological charge $\nu=(Q_{x},w)$.

The first equality in Eq.~(\ref{eq:BBCmin}) states
that we will find $N_{c}=4|w|$ corner states on the Fermi surface
of the 2D system at $2/3$ and $1/3$ fillings provided that $Q_{x}=\pm i$
and $\pm k$, respectively. In these two cases, the subsystem $H_{y}$
owns two Abelian edge states $|\varphi_{0}^{(1)}\rangle$ and $|\varphi_{0}^{(2)}\rangle$
at zero energy for $w=1$, while the subsystem $H_{x}$ has two non-Abelian
edge states $|\psi_{E}^{(1)}\rangle$ and $|\psi_{E}^{(2)}\rangle$
at the same energy $E\in\Delta_{23}$ ($E\in\Delta_{12}$) for $Q_{x}=\pm i$
($Q_{x}=\pm k$), where $\Delta_{12}$ ($\Delta_{23}$) denotes the spectral
gap between the first and second (second and third) bulk bands of
$H_{x}$. The hybridization of these edge states along two orthogonal
dimensions results in four corner states $|\Psi_{E}^{(\ell,\ell')}\rangle=|\psi_{E}^{(\ell)}\rangle\otimes|\varphi_{0}^{(\ell')}\rangle$
at the same energy $E$ of the 2D Hamiltonian $H=H_{x}\oplus H_{y}$,
where $\ell,\ell'=1,2$. Their fourfold degeneracy is protected by
the combined ${\cal PT}$ and chiral symmetries of subsystems $H_{x}$
and $H_{y}$.

The second equality in Eq.~(\ref{eq:BBCmin}) states
that we have $N_{c}=4|w|$ corner states on the Fermi surface
of the 2D system at both $1/3$ and $2/3$ fillings if $Q_{x}=\pm j$
or $-1$. In these two cases, the subsystem $H_{y}$ again has two
Abelian edge states $|\varphi_{0}^{(1)}\rangle$ and $|\varphi_{0}^{(2)}\rangle$
at zero energy for $w=1$. Meanwhile, the subsystem $H_{x}$ has two
edge states $|\psi_{E}^{(1)}\rangle$ and $|\psi_{E}^{(2)}\rangle$
at energy $E\in\Delta_{12}$ and two other edge states
$|\psi_{E'}^{(1)}\rangle$ and $|\psi_{E'}^{(2)}\rangle$ at energy
$E'\in\Delta_{23}$. Their hybridization along two
orthogonal dimensions leads to two groups of fourfold degenerate
corner states characterized by NATCs at the energies $E$ and $E'$, given by $|\Psi_{E}^{(\ell,\ell')}\rangle=|\psi_{E}^{(\ell)}\rangle\otimes|\varphi_{0}^{(\ell')}\rangle$
and $|\Psi_{E'}^{(\ell,\ell')}\rangle=|\psi_{E'}^{(\ell)}\rangle\otimes|\varphi_{0}^{(\ell')}\rangle$
for $\ell,\ell'=1,2$. The degeneracy within each group of these corner
states is also jointly protected by the ${\cal PT}$ and chiral symmetries
of subsystems $H_{x}$ and $H_{y}$.

The third equality in Eq.~(\ref{eq:BBCmin}) states
that we have no corner states anywhere if $Q_{x}=1$, regardless
of the value of winding number $w$. In this case, the $H_{x}$
does not support edge states associated with NATCs. The edge zero modes of $H_{y}$,
if exist, can only couple to the extended bulk states of $H_{x}$,
yielding delocalized states in the 2D bulk of the full system
$H$ or along its 1D edges.

The above analyses confirm that the number of corner
states in each of our SOTPs is determined
by the hybridized NATC $\nu=(Q_{x},w)$.
Yet, as the edge states of subsystem $H_{y}$ are pinned to zero energy,
the energy spans of our corner states are controlled by the spectrum
of edge states of subsystem $H_{x}$ under OBC. As a remark,
one may not be able to deduce whether the bulk charge is $(j,1)$
or $(-1,1)$ by directly viewing the spectrum of $H$ under OBC, as
the corner states look similar in these two cases.
This issue might be resolved by comparing the following two situations.
Let us assume that two distinct phases of Hamiltonian $H$, with
bulk charges $\nu$ and $\nu'=(i,1)$, are realized
in two square lattices. The two systems are then coupled together
at a single corner. If the invariant $\nu=(j,1)$, the quotient rule
of NATCs will induce a domain-wall charge $\nu/\nu'=(j/i,1)=(k,1)$,
yielding corner states at the coupled corner when the composite system
$H$ is set at $1/3$ filling. On the other hand, if $\nu=(-1,1)$,
the quotient rule of NATCs will induce a domain-wall charge $\nu/\nu'=(-1/i,1)=(i,1)$,
yielding corner states at the coupled corner when the composite system
$H$ is set at $2/3$ filling. The non-Abelian SOTPs with bulk charges
$(j,1)$ and $(-1,1)$ can then be distinguished by investigating
their domain-wall behaviors under OBC.

To illustrate the bulk-corner correspondence,
we computed the spectra and corner states of our minimal model
$H$ for its typical non-Abelian SOTPs.
The results are shown in Fig.~\ref{fig: band and corner}. To distinguish the bulk
and corner states in the spectra, we evaluate the IPR
for each 2D eigenstate of $H$ in the lattice space,
which is given by
\begin{equation}
{\rm IPR}_{q}=\sum_{n=1}^{N}\sum_{m=1}^{M/2}\sum_{\alpha=A,B,C}|\langle n,m,\alpha|\Psi_{q}\rangle|^{4}.\label{eq:IPRq}
\end{equation}
Here, $|\Psi_{q}\rangle$ denotes the $q$th eigenstate of $H$. The
total number of such states is $3MN$. If $|\Psi_{q}\rangle$
represents an extended bulk state, we will have ${\rm IPR}_{q}\rightarrow0$
in the limits $N,M\rightarrow\infty$, and the numerical value of
${\rm IPR}_{q}$ would become vanishingly small when the total number
of sites $N\times M$ is large enough. If $|\Psi_{q}\rangle$
describes a localized corner state, we will obtain a finite ${\rm IPR}_{q}$
even in the limits $N,M\rightarrow\infty$, and the numerical value
of ${\rm IPR}_{q}$ would be drastically different from zero for finite
number of sites $N\times M$. Using the value of ${\rm IPR}_{q}$
as the color code of each eigenstate then allows us to discriminate
bulk and corner states in the spectrum from their localization nature.

In Fig.~\ref{fig: band and corner}(a), we observe four degenerate eigenmodes
in the gapped energy regime $E\in\Delta_{23}$. They are indeed
localized at the four corners of the 2D lattice, as reflected by their
probability distributions in Fig.~\ref{fig: band and corner}(e). For this example, the system
parameters are taken in the region with topological charge $\nu=(i,1)$
in Fig.~\ref{fig: Phase_Diagram}. The hybridized non-Abelian bulk-corner correspondence,
as described by the first equality in Eq.~(\ref{eq:BBCmin}), is then
verified. In Fig.~\ref{fig: band and corner}(b), we observe four degenerate eigenmodes in each
of the energy regions $E\in\Delta_{12}$ and $E'\in\Delta_{23}$.
These eight eigenstates are all localized at the corners of the 2D
lattice, as shown in Fig.~\ref{fig: band and corner}(f). In this case, the system parameters
are chosen in the regime with topological charge $\nu=(j,1)$ in Fig.~\ref{fig: Phase_Diagram}.
The bulk-corner correspondence, as described by the second equality
in Eq.~(\ref{eq:BBCmin}), is again verified. In Fig.~\ref{fig: band and corner}(c), we find
four degenerate eigenstates in the gapped energy regime $E\in\Delta_{12}$.
They are located at the four corners of the 2D lattice, as presented
in Fig.~\ref{fig: band and corner}(g). Moreover, the system parameters for this example are
chosen in the region with topological charge $\nu=(k,1)$ in 
Fig.~\ref{fig: Phase_Diagram}, thus confirming the bulk-corner correspondence in the first equality of Eq.~(\ref{eq:BBCmin}). 
Finally, we identify four degenerate eigenstates in each of the energy regions $E\in\Delta_{12}$ and $E'\in\Delta_{23}$ in Fig.~\ref{fig: band and corner}(d).
Their spatial profiles are localized at the four corners of the 2D lattice in Fig.~\ref{fig: band and corner}(h). In this case, the system parameters
are chosen in the regime with topological charge $\nu=(-1,1)$ in
Fig.~\ref{fig: Phase_Diagram}, verifying the bulk-corner correspondence in the second equality
of Eq.~(\ref{eq:BBCmin}). 
Spectrally, these corner states can overlap with and become embedded in the bulk energy bands. Such states may be viewed as bound states in the continuum (BICs), which lie within the bulk energy range while remaining spatially localized and effectively decoupled from extended states. The robustness of these BICs is ensured by the topological invariants of the constituent subsystems, such that the corner modes remain stable unless a global topological phase transition occurs. This observation further supports our classification of the system into SOTPs, in which strict spectral gaps are not required for the topological protection. Further details are given in Appendix~\ref{app: disorder}.
In summary, we have checked the bulk-corner
correspondence in Eq.~(\ref{eq:BBCmin}) for all typical phases of
our 2D non-Abelian SOTPs. Compared to Abelian cases, the key
difference is that due to the quaternion component $Q_{x}$
in the hybridized topological charge $\nu$, each of the phases
and its corner modes here inherit non-Abelian characteristics.

In more general situations, since the hybridized NATC $(-1,1)=(i^{2},1)=(j^{2},1)=(k^{2},1)$, 2D non-Abelian
SOTPs may also possess corner modes exclusively in spectral ranges
$\Delta_{12}$ or $\Delta_{23}$, yielding richer configurations of
corner states associated to NATCs. We discuss explicit examples for these
cases in Appendix \ref{app: another -1}. Besides, 1D edge bands of non-Abelian origin
can arise in our system if the quaternion component $Q_{x}$
of $\nu$ is nontrivial, as will be uncovered in the next subsection.

\subsection{Non-Abelian topological edge bands}\label{sec: y trivial}

\begin{figure*}
    \centering
    \includegraphics[width=1\linewidth]{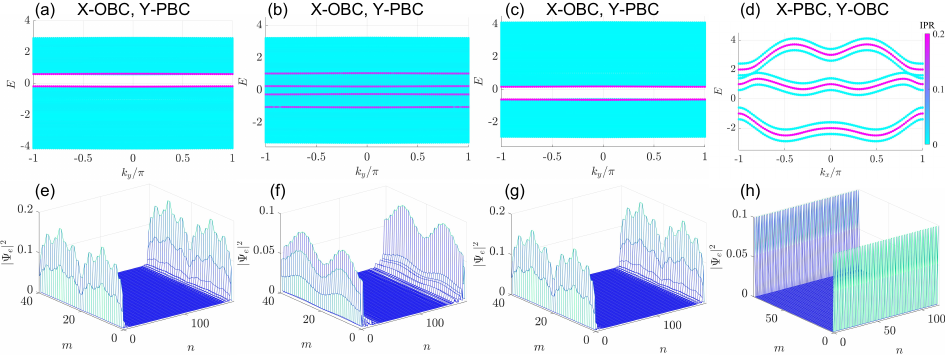}
    \caption{Spectrum and edge states of the minimal non-Abelian model of SOTPs. (a)--(d) show the energy spectra. For (a)--(c), we take the OBC (PBC) along the $x$ ($y$) direction of the lattice. For (d), we take PBC (OBC) along the $x$ ($y$) direction of the lattice. In (e)--(h), we plot edge states with the largest twenty IPRs in (a)--(d). System parameters used in (a)--(c) are identical to those in Figs.~\ref{fig: band and corner}(a)--\ref{fig: band and corner}(c). (c) and (d) share the same parameters but have different boundary conditions.}\label{fig: edge state 2D}
\end{figure*}

As mentioned in the last subsection, if any one of the
components of $\nu=(Q_{x},w)$ is nontrivial, the edge states
of its associated subsystem can bind to the bulk states of the rest
subsystem, forming boundaries states localized along a single direction
at certain edges of the 2D lattice. Since the formation of these edge
states depends on the type of boundary conditions, we refer to them
as weak topological edge states and group them into two classes depending
on which component of $\nu$ is nontrivial. Interestingly, these edge
states could also be of non-Abelian origins when they are topological,
as clarified below.

In the first class, the hybridized charge of the 2D
system takes the form $\nu=(Q_{x},0)$ with $Q_{x}\neq1$. In this
case, the subsystem $H_{y}$ hosts only trivial insulator phases.
Taking the OBC and PBC separately along $x$ and $y$ directions,
we can obtain a set of edge bands vs the conserved quasimomentum $k_{y}$
of $H_{y}(k_{y})$, as illustrated in Figs.~\ref{fig: edge state 2D}(a)--\ref{fig: edge state 2D}(c).
For each state $|\Psi(k_{y})\rangle$, the IPR there is defined as
$\sum_{n=1}^{N}\sum_{\alpha=A,B,C}|\langle n,\alpha|\Psi(k_{y})\rangle|^{4}$.
The total number of these edge bands $N_{ex}$ is determined by the
quaternion charge $Q_{x}$. For our minimal model, the resulting bulk-edge
correspondence reads
\begin{equation}
N_{ex}=\begin{cases}
4|Q_{x}^{2}|, & Q_{x}=\pm i,\pm k,\\
8|Q_{x}^{2}|, & Q_{x}=\pm j,-1.
\end{cases}\label{eq:Nex}
\end{equation}
These relations are of non-Abelian origin due to the noncommutative
nature of $Q_{x}$. We will be left with $N_{ex}=0$ if $Q_{x}=1$.
The prefactors of $|Q_{x}^{2}|$ come from the fact that each non-Abelian
edge state of $H_{x}$ is twofold degenerate and $H_{y}(k_{y})$
has two bulk bands at $\pm E_{y}(k_{y})$. Therefore, if $Q_{x}=\pm i$
($\pm k$), the two edge states $|\psi_{E}^{(1)}\rangle$ and $|\psi_{E}^{(2)}\rangle$
of $H_{x}$ will couple to the Bloch states $|\varphi_{\pm}(k_{y})\rangle$
of $H_{y}(k_{y})$ with energies $\pm E_{y}(k_{y})$ at each $k_{y}$,
yielding two sets of twofold degenerate edge bands $|\Psi_{E,\pm}^{(\ell)}(k_{y})\rangle=|\psi_{E}^{(\ell)}\rangle\otimes|\varphi_{\pm}(k_{y})\rangle$
with $\ell=1,2$ and $k_{y}\in[-\pi,\pi)$ at the energies $E\pm E_{y}(k_{y})$,
where the $E\in\Delta_{23}$ ($E\in\Delta_{12}$) for $H_{x}$. Meanwhile,
if $Q_{x}=\pm j$ or $-1$, the four edge states $|\psi_{E}^{(1)}\rangle$,
$|\psi_{E}^{(2)}\rangle$, $|\psi_{E'}^{(1)}\rangle$ and $|\psi_{E'}^{(2)}\rangle$
of $H_{x}$ will bind to the Bloch states $|\varphi_{\pm}(k_{y})\rangle$
of $H_{y}(k_{y})$ with energies $\pm E_{y}(k_{y})$ at each $k_{y}$,
forming four sets of twofold-degenerate edge bands $|\Psi_{E,\pm}^{(\ell)}(k_{y})\rangle=|\psi_{E}^{(\ell)}\rangle\otimes|\varphi_{\pm}(k_{y})\rangle$
and $|\Psi_{E',\pm}^{(\ell)}(k_{y})\rangle=|\psi_{E'}^{(\ell)}\rangle\otimes|\varphi_{\pm}(k_{y})\rangle$
with $\ell=1,2$ and $k_{y}\in[-\pi,\pi)$ at the energies $E\pm E_{y}(k_{y})$
and $E'\pm E_{y}(k_{y})$, where the $E\in\Delta_{12}$ and $E'\in\Delta_{23}$
for $H_{x}$. As the topology of these edge bands is characterized
by the quaternion charge $Q_{x}$, they constitute a unique type of
non-Abelian weak topological edge states protected
by the ${\cal PT}$ symmetry of $H_{x}$, which implies that they
could even be robust to perturbations that break the chiral symmetry
of subsystem $H_{y}$. In this sense, we may regard this class
of edge bands as sub-symmetry-protected non-Abelian topological states.
In Figs.~\ref{fig: edge state 2D}(e)--\ref{fig: edge state 2D}(g), we show the edge states in Figs.~\ref{fig: edge state 2D}(a)--\ref{fig: edge state 2D}(c)
under the OBC along both dimensions. The observed edge states
are indeed localized only along $x$ direction at the left and right
boundaries of the 2D lattice.

In the second class, the hybridized charge of the
2D system reads $\nu=(Q_{x},1)$, such that the subsystem $H_{y}$
hosts topological insulator phases with two edge zero modes under
OBC. Taking the PBC and OBC separately along $x$ and $y$ directions,
we obtain a set of edge bands vs the conserved quasimomentum $k_{x}$
of $H_{x}(k_{x})$, as illustrated in Fig.~\ref{fig: edge state 2D}(d). There,
the IPR is defined as $\sum_{m=1}^{M/2}|\langle m|\Psi(k_{x})\rangle|^{4}$
for each state $|\Psi(k_{x})\rangle$. For any $Q_{x}$,
the total number of such edge bands is given by
\begin{equation}
N_{ey}=6|w|,\qquad Q_{x}\in\mathbb{Q}_{8}.\label{eq:Ney}
\end{equation}
The prefactor ``$6$'' comes from the fact that the subsystem $H_{x}(k_{x})$
holds three bulk bands $\{E_{p}(k_{x})|p=1,2,3\}$ with Bloch states
$\{|\psi_{p}(k_{x})\rangle|p=1,2,3\}$ for $k_{x}\in[-\pi,\pi)$,
regardless of its quaternion charge, and each edge state of $H_{y}$
is twofold degenerate at zero energy. For our minimal model with $w=1$,
the resulting six edge bands form three degenerate subsets with dispersion
relations $E_{p}(k_{x})$ and eigenstates
$|\Psi_{p}^{(1)}(k_{x})\rangle=|\psi_{p}(k_{x})\rangle\otimes|\varphi_{0}^{(1)}\rangle$,
$|\Psi_{p}^{(2)}(k_{x})\rangle=|\psi_{p}(k_{x})\rangle\otimes|\varphi_{0}^{(2)}\rangle$
for $p=1,2,3$, where $|\varphi_{0}^{(1)}\rangle$ and $|\varphi_{0}^{(2)}\rangle$
are edge zero modes of $H_{y}$. In each subset, the degeneracy
of edge bands is protected by the chiral symmetry of $H_{y}$, which
is then robust to perturbations that break the ${\cal PT}$ symmetry
of subsystem $H_{x}(k_{x})$. In this sense, we can regard this class
of edge bands as sub-symmetry-protected topological states, which
are formally of Abelian origin as their numbers are solely controlled
by the winding number $w$. Nevertheless, when the ${\cal PT}$ symmetry
is preserved and the quaternion charge $Q_{x}\neq1$, nontrivial eigenframe
rotations could develop within the subspace of edge bands of $H(k_{x})=H_{x}(k_{x})\oplus H_{y}$,
yielding weak topological edge states with non-Abelian characters,
as revealed by the nontrivial value of $Q_{x}$. The manipulation
of these non-Abelian states at the edges may further lead to ${\cal PT}$-symmetry-protected
qubit operations, which may find applications in future quantum
technologies. Finally, we confirm in Fig.~\ref{fig: edge state 2D}(h) that the edge states
of our second class are indeed localized only along $y$ direction
at the top and bottom boundaries of the 2D lattice when OBC is taken
along both dimensions.

In summary, both the edge and corner states of our
system are characterized by the hybridized bulk topological charge
$\nu=(Q_{x},w)$. Non-Abelian second-order corner states arise if
and only if both components of $\nu$ are nontrivial and can coexist
with first-order topological edge states of non-Abelian characters.
If merely one component of $\nu$ is nontrivial, the system
supports weak topological edge states of either non-Abelian (with $w=0$)
or Abelian (with $Q_{x}=1$) origins, which are protected by the sub-symmetry
of the rest nontrivial subsystem. When both components of $\nu$ are
trivial, the system becomes a trivial insulator, hosting neither edge
nor corner topological states. We thus established a complete
bulk-edge-corner correspondence for 2D non-Abelian HOTPs satisfying our
coupled-wire construction.
It deserves to emphasize that even though we focused on a detailed study of
only the minimal 2D model, our coupled-wire scheme
can be used to construct non-Abelian HOTPs of any order in any spatial dimensions without
great efforts.

\section{Conclusion and discussion}\label{sec: conclusion}
In this work, we introduced a theoretical framework to construct pristine and hybridized non-Abelian HOTPs. Their underlying topology was found to be characterized by the tensor product of topological charges coming from distinct Abelian and non-Abelian groups of constituent subsystems. Focusing on the minimal model of non-Abelian HOTPs as an illustrative example, we revealed a rich set of non-Abelian SOTPs in experimentally realizable 2D lattices, which were protected by the united chiral and ${\cal PT}$ symmetries. Each of the discovered phases was uniquely identified by a two-component change $\nu=(Q_{x},w)$, with $Q_{x}$ and $w$ be quaternions and integers, respectively. A phase is then topological if and only if both two components of $\nu$ are nontrivial. Despite uncovering a series of non-Abelian SOTPs and phase transitions, we also found the correspondence between the bulk topological invariant $\nu$, the configurations of corner states in rectangular geometries and of edge states in cylinder geometries for each of the non-Abelian SOTPs, thereby establishing a unified perspective on their bulk-edge-corner correspondences. Notably, we obtained 1D weak topological edge bands of non-Abelian origins, whose topologically-protected edge states may find applications in quantum information and computation related tasks. Specifically, these boundary modes may serve as high-fidelity channels for robust quantum state transfer in superconducting qubit chains~\cite{QST}, effectively protecting quantum information from local disorder and decoherence. Furthermore, the topological degeneracy of these states enables them to act as coherent mediators for long-distance qubit interactions~\cite{QIM}, facilitating scalable entanglement and quantum logic operations in waveguide QED or photonic platforms~\cite{QED,SQbit}. Distinct from conventional Abelian setups, our multi-gap models characterized by hybridized NATCs offer richer boundary modes distributed across various gaps or bulk spectra. By tuning the coupling strengths, one can flexibly steer the non-Abelian topological phase transitions to obtain optimal edge or corner states for specific quantum tasks.
Overall, our work not only extends the study of HOTPs to non-Abelian domains, but also blends the Abelian and non-Abelian topological phases into an individual context, opening avenues for more systematic classifications of their topological nature.

In experiments, transmission line networks have been proven to be efficient in simulating high-dimensional lattice models and probing their physical properties. For example, three dimensional waveguide networks have been employed to observe Anderson localization~\cite{zhang1998observation}, while 2D honeycomb networks have been used to emulate Chern insulators~\cite{jiang2019experimental}. More importantly, this platform has been applied to the realization of non-Abelian first-order topological phases by connecting nodes through waveguides~\cite{NATI1,NATI2}. One can then construct quasi-1D lattice models with long-range hoppings, where each sublattice site within a unit cell serves as a node and the intercell couplings are implemented as long-range connections.
When all nodes are connected by cables of equal length and each node is attached to the same number of cables, the wave equation at each node in the lossless limit becomes equivalent to the Schr\"odinger equation of a tight-binding model. In this situation, with all cables having identical lengths, different coupling strengths are implemented by varying the number of connections between nodes. Onsite potentials can be realized by adding self-connected cables to the nodes. Under these conditions, a transmission line network provides a flexible setup to simulate our lattice models. The measurement of voltage at each node corresponds to probing the wave amplitude. By injecting a signal at one end and measuring the resulting global voltage distribution, one can directly obtain the spatial profile of the wave function.
These advances provide strong experimental support for the feasibility of realizing our model in transmission line networks. Therefore, we expect the full topological features of our model, including the non-Abelian corner and edge states, to be observable in such setups under current experimental capabilities. 
In transmission line networks with periodic structures, the Bloch eigenvectors can be extracted by Fourier transforming the voltage vectors at real-space nodes. This enables reconstructions of the momentum evolution of eigenstates in $k$ space, thereby identifying the related topological charge.

In future work, since our coupled-wire scheme is generic, it can be used to construct pristine non-Abelian SOTPs in two dimensions and other non-Abelian HOTPs in higher dimensions, whose topological charges are still obtainable following our theoretical framework. Beyond equilibrium, unique non-Abelian HOTPs may arise in non-Hermitian and Floquet systems, whose topological properties are still awaited to be revealed. Finally, the robustness of non-Abelian HOTPs to disorder and interactions would be important for their practical applications, which thus deserve more thorough explorations.

\begin{acknowledgments}
This work is supported by the National Natural Science Foundation of China (Grants No.~12275260 and No.~11905211), the Fundamental Research Funds for the Central Universities (Grant No.~202364008), and the Young Talents Project of Ocean University of China.
\end{acknowledgments}

\appendix

\section{Zak phase}\label{app: Zak phase}

In this Appendix, we analyze the topological properties of our system from the perspective of Abelian geometric phase. We reveal that this approach becomes insufficient for depicting non-Abelian HOTPs~\cite{NATS1,NATI1}.

In 1D systems, the Zak phase \cite{Zak1989} offers a geometric indicator of topological states. 
A quantized Zak phase $\gamma=\pi\mod 2\pi)$ typically signals a nontrivial phase that supports degenerate edge states, whereas $\gamma=0$ corresponds to a topologically trivial insulating phase.
For an isolated band, the Zak phase is given by the integral of Berry connection $A(k)$ over the Brillouin zone, i.e,
\begin{equation}
\gamma = \oint_{\text{BZ}} A(k) \, dk = i \oint_{\text{BZ}} \braket{ u_k | \partial_k | u_k} dk,  \label{eq: Zak phase}
\end{equation}
where $\ket{u_k}$ denotes the Bloch eigenstate. Therefore, the Zak phase corresponds to the geometric Berry phase accumulated by a Bloch state during an adiabatic traversal of the Brillouin zone.

For the SSH model, the Zak phase serves as a topological invariant equivalent to the winding number, which 
could faithfully characterize the bulk-edge correspondence.
For our 1D Hamiltonian $H_y(k_{y})$, the Berry connection of its valence band is given by
\begin{equation}
A(k_{y}) =i \braket{\psi_{-}(k_{y}) | \partial_{k_{y}} | \psi_{-}(k_{y})}, \label{eq: y_Zak phase}
\end{equation}
where $\ket{\psi_{-}(k_{y})}$ denotes the eigenstate of lower band, yielding the Zak phase following Eq.~(\ref{eq: Zak phase}). 
In the main text, Fig.~\ref{fig: 1D spectrum}(a) shows the spectrum of SSH model as our subsystem $H_y$. 
Using Eqs.~(\ref{eq: Zak phase}) and (\ref{eq: y_Zak phase}), we find the corresponding Zak phase in Fig.~\ref{fig: 1D spectrum}(a).
When $H_{y}$ lies in a topological phase, the Zak phase is $\gamma_{y} = \pi$, and a pair of edge states arise in the spectrum at $E_{y}=0$. 
When $\gamma_{y} = 0$, there are no edge states and the system becomes topologically trivial.

However, the Zak phase is no longer a reliable topological invariant for our subsystem $H_x$.
To see this, let us try to use the Zak phase of occupied bands as an indicator of edge states within each gap.
For the lower gap, we focus on the Zak phase $\gamma_{12}=\gamma_{1}$ of the first band of $H_x(k_{x})$.
For the upper gap, we use the sum of Zak phases of the first and second bands, given by $\gamma_{23}=\gamma_{1}+\gamma_{2}$.
For our three-band Hamiltonian $H_x(k_{x})$, we have $\gamma_{x}\equiv(\gamma_{12},\gamma_{23})$, and the Zak phase of each band reads
\begin{equation}
    \gamma_{p}=i\oint \braket{\psi_{p}(k_{x}) | \partial_{k_{x}} | \psi_{p}(k_{x}) } \, dk_x,\quad p=1,2,3,   \label{x_Zak phase}
\end{equation}
where $\ket{\psi_{p}(k_{x})}$ is the $p$th eigenvector of $H_x(k_{x})$.
In Figs.~\ref{fig: 1D spectrum}(b)--\ref{fig: 1D spectrum}(e), we show the energy spectra and IPR of $H_{x}$ under OBC, with the Zak phases labeled in each panel.
For NATCs $Q_{x}=i,j,k,1$, the Zak phase correctly reflects the topology of each band gap. 
For example, in Fig.~\ref{fig: 1D spectrum}(b), a pair of edge states appears in the upper band gap, corresponding to $\gamma_{x}=(0,\pi)$ and NATC $i$. 
In Fig.~\ref{fig: 1D spectrum}(d), two pairs of edge states arise in both gaps, corresponding to $\gamma_{x}=(\pi,\pi)$ and NATC $j$.
In Fig.~\ref{fig: 1D spectrum}(e), a pair of edge states appears in the lower band gap, corresponding to $\gamma_{x}=(\pi,0)$ and NATC $k$.
For NATC $-1$, we find two pairs of edge states under OBC.
However, the Zak phase is $\gamma_{x}=(0,0)$ in this case, identical to the trivial phase $Q_{x}=1$ without edge states. 
This contradiction shows that the Zak phase fails to capture all phases of our model, making the introduction of NATCs essential.
From another viewpoint, the eigenframe rotations of $H_x$ show at least five distinct patterns [Figs.~\ref{fig: Topological geometry}(b)--\ref{fig: Topological geometry}(e)].
However, a system with two band gaps allows at most $2^{2}=4$ combinations of Zak phases. 
This discrepancy demonstrates that the Zak phase cannot capture the overall non-Abelian topology encoded in $H_x$.

\section{\texorpdfstring{Non-Abelian SOTPs with charge $(-1,1)$}{Title with math}}\label{app: another -1}
In Sec.~\ref{sec: corner states}, we examined the corner states arising in various topological phases of Fig.~\ref{fig: Phase_Diagram}. 
In this Appendix, we focus on a particular phase with charge $\nu=(-1,1)$ and analyze its distinctive features in detail.
As discussed in Sec.~\ref{sec: topological invariant}, the topological phase of $H_x$ with $Q_{x}=-1$ has no analogs in Zak-phase descriptions.
Under OBC, this phase further reveals enriched edge state configurations, which distinguish it from both Abelian and other non-Abelian topological phases.
Due to quaternion relations $i^{2}=j^{2}=k^{2}=-1$, the case $Q_{x}=-1$ may be realized by doubled eigenframe rotations of the charge $i$, $j$ or $k$. 
Although these quaternion elements are equivalent and can be continuously deformed into one another~\cite{NATI1}, they exhibit distinct edge state structures under OBC.

\begin{figure}
    \centering
    \includegraphics[width=1\linewidth]{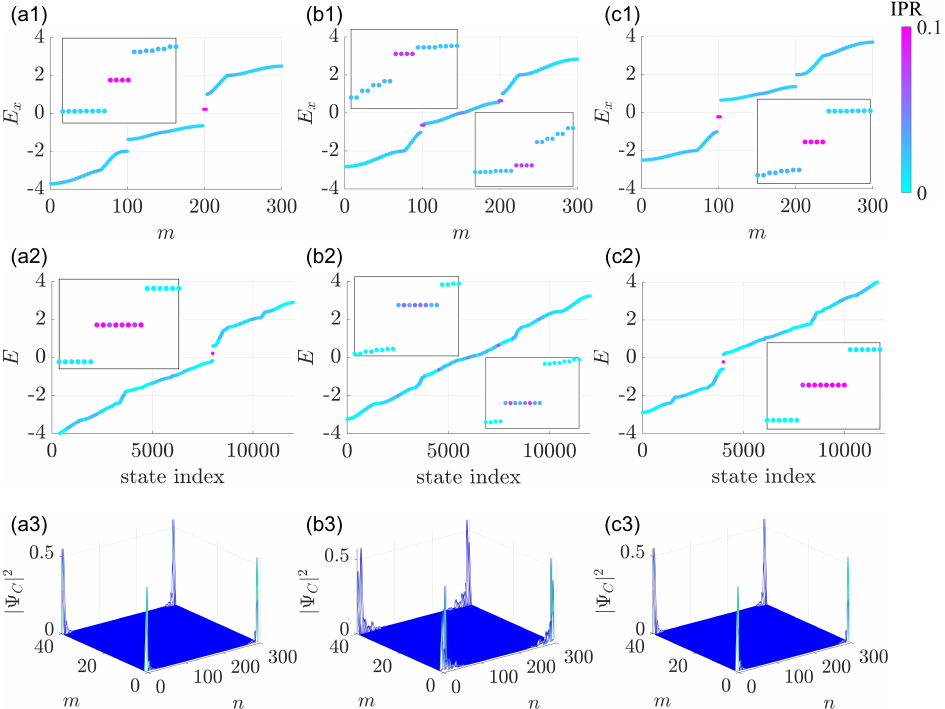}
    \caption{Spectra and IPR of non-Abelian SOTPs under OBC. (a1)--(c1) show energy spectra of the Hamiltonian $H'$. They share the NATC $Q_{x}=-1$. (a2)--(c2) show energy spectra of the 2D Hamiltonian $\mathcal{H}$. (a3)--(c3) show probability distributions of the corner states in (a2)--(c2). The insets highlight the edge and corner states. (a1)--(c1) and (a2)--(c2) have the same color bar, which represents the magnitude of IPR. In (a)--(c), the NNN hopping amplitudes are the same as the NN hopping amplitudes in Figs.~\ref{fig: band and corner}(a)--\ref{fig: band and corner}(c), i.e., $J'_{\alpha\alpha'}=J_{\alpha\alpha'}$.}\label{fig: other -1}
\end{figure}

To further reveal the peculiarity of the $(-1,1)$ phase, we introduce next-nearest-neighbor (NNN) hoppings along $x$ direction in the model shown in Fig.~\ref{fig: lattice model}. The NNN Hamiltonian in momentum space reads
\begin{equation}
        H'_x(k_{x})=
        \begin{pmatrix}
        V'_{AA} & 2u'\sin{2k_{x}} & 0\\
        2u'\sin{2k_{x}} & V'_{BB} & 2v'\sin{2k_{x}}\\
        0 & 2v'\sin{2k_{x}} & V'_{CC}
        \end{pmatrix},\\
\end{equation}
where $V'_{\alpha\alpha}\equiv S_\alpha+2J'_{\alpha\alpha}\cos{2k_x}$ for $\alpha=A,B,C$.
The spectra of $H'_x(k_{x})$ under OBC are shown in Figs.~\ref{fig: other -1}(a1)--\ref{fig: other -1}(c1).
Although sharing the same NATC $Q_{x}=-1$, these cases exhibit different edge-mode distributions.
This discrepancy comes from different physical effects associated with $i^{2}$, $j^{2}$ and $j^{2}$.
In particular, the spectrum in Fig.~\ref{fig: other -1}(a1) hosts a single set of fourfold degenerate edge states in the upper band gap, whereas the spectrum in Fig.~\ref{fig: other -1}(b1) contains two such sets, each residing in a different band gap. Thus, even though $i^{2}$ and $j^{2}$ are equivalent representatives of $Q_x=-1$, the physical arrangements of edge states retain clear signatures of their distinctive eigenframe-rotation axes.
Note in passing that all phases with $Q_x=-1$ have a trivial Zak phase $\gamma_{x}=(0,0)$.

The 1D non-Abelian phase with $Q_x = -1$ in different quaternion representatives shows distinct edge state configurations under OBC, yielding different degeneracies and spectral locations for corner states in the 2D system.
With NNN hoppings, the 2D model has the Hamiltonian $\mathcal{H}({\bf k})=H'_x(k_{x})\otimes\sigma_{0}+{\mathbb I}_{3}\otimes H_y(k_{y})$. Its spectra, corresponding to the same hybridized charge $\nu=(-1,1)$, are shown in Figs.~\ref{fig: other -1}(a2)--\ref{fig: other -1}(c2).
In each case, the corner modes arise from the interplay between topological edge states contributed by $H'_x$ and $H_y$.
In Figs.~\ref{fig: other -1}(a2) and \ref{fig: other -1}(c2), there is a set of corner states with eightfold degeneracy.
Their energies correspond to those of edge states in Figs.~\ref{fig: other -1}(a1) and \ref{fig: other -1}(c1).
In Fig.~\ref{fig: other -1}(b2), there are two sets of eightfold degenerate corner states.
Their energies are consistent with those of edge states in Fig.~\ref{fig: other -1}(b1).
The probability distributions of localized states in Figs.~\ref{fig: other -1}(a2)--\ref{fig: other -1}(c2) are shown in Figs.~\ref{fig: other -1}(a3)--\ref{fig: other -1}(c3), 
respectively, which are indeed pinned to the system corners.
Here, the key feature of the non-Abelian SOTPs with charge $\nu=(-1,1)$ is that while the bulk invariant determines the overall topological class, the corner states can display distinct configurations. That is, although the $Q_x=-1$ charge generated by different eigenframe rotations is topologically equivalent in the bulk, they leave distinct signatures in the energy spectra under OBC. These corner state distributions can be smoothly transformed into one another without closing band gaps.
Essentially, the eigenenergies of our corner states are governed by the quaternion component $Q_{x}$, while the winding number $w$ leads to
an additional twofold degeneracy for each corner state,
yielding the final configuration of hybridized corner states arising from both Abelian and non-Abelian topological charges.

\section{Robustness of corner states as BICs}\label{app: disorder}

\begin{figure}
    \centering
    \includegraphics[width=0.9\linewidth]{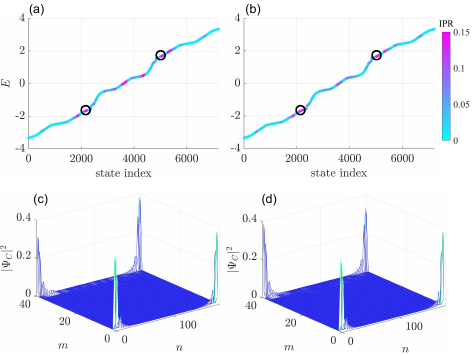}
    \caption{Spectrum and corner states of the minimal 2D non-Abelian model with disorder $\lambda \epsilon_{nm\alpha}$ under OBC along both dimensions. (a) and (b) show the energy spectra for two independent realizations of the random disorder. Corner states are highlighted by black circles.
    The shared color bar indicates the IPR of each eigenstate. (c) and (d) show the probability distributions of corner states in (a) and (b). System parameters are same with Fig.~\ref{fig: band and corner}(d). The disorder strength $\lambda=0.2$.}\label{fig: disorder}
\end{figure}

In this Appendix, we demonstrate the stability of topological corner states in the absence of band gaps in our system. Specifically, by introducing disorder to the onsite potential, we verify that the corner states remain robust under small perturbations.

The Kronecker sum structure implies that our 2D model is not necessarily gapped. In particular, BICs can arise as shown in Fig.~\ref{fig: band and corner}(d). Nevertheless, according to the construction of our model, the emergence of corner states is determined by the topological properties of the constituent subsystems. As a result, the presence or absence of a global energy gap does not affect the existence of the topological corner states, so long as a direct band touching has not been developed. This conclusion is further supported by our numerical simulations. We introduce onsite disorder of the form $\lambda \epsilon_{nm\alpha}$, where $\alpha = A, B, C$, i.e., let $S_{\alpha} \rightarrow S_{\alpha}+\lambda \epsilon_{nm\alpha}$. $\lambda$ denotes the disorder strength, and $\epsilon_{nm\alpha} \in [-1,1]$ is a uniformly distributed random variable for all $n$, $m$, and $\alpha$. We have checked the energy spectra over a large number ($\simeq100$) of disorder realizations, from which two representative cases are selected and shown in Fig.~\ref{fig: disorder}(a) and \ref{fig: disorder}(b). The introduction of disorder leads to the emergence of additional localized states, while the corner states are highlighted by black circles. Although disorder induces small shifts in the eigenenergies of the corner states, their probability distributions remain localized at the corners, as shown in Fig.~\ref{fig: disorder}(c) and \ref{fig: disorder}(d). This demonstrates that weak disorder does not destroy the topological corner states in our system. Therefore, even in the absence of bulk band gaps, the corner states associated to NATCs remain robust and well-defined.

\bibliography{reference.bib}

\end{document}